\RequirePackage{fix-cm}
\documentclass[twocolumn,epjc3]{svjour3}  
\smartqed  
\RequirePackage{graphicx,graphics,amsmath}
\usepackage{xcolor}
\usepackage{float}
\RequirePackage{latexsym}
\usepackage[utf8x]{inputenc}
\journalname{Eur. Phys. J. C}
\begin{document}

\title{Tidal effect on the gyroscopic precession around a compact star}



\author{Kamal Krishna Nath\thanksref{e1,addr1}
	   Debojoti Kuzur\thanksref{e2,addr1}
        \and
        Ritam Mallick\thanksref{e3,addr1} 
}

\thankstext{e1}{e-mail: knath@iiserb.ac.in}
\thankstext{e2}{e-mail: debojoti16@iiserb.ac.in}
\thankstext{e3}{e-mail: mallick@iiserb.ac.in}

\institute{Indian Institute of Science Education and Research Bhopal, Bhopal, India \label{addr1}
}

\date{Received: date / Accepted: date}

\maketitle

\begin{abstract}
General relativistic effects in the spacetime around the massive astrophysical objects can be captured using a spinning test gyro orbiting around the object in a circular geodesic. This article discusses how the tidal disruption due to a companion object affects the precession frequency of a spinning gyro orbiting around a compact astrophysical object. The precession frequency is studied in a region of space around the central object using a perturbative approach. In this study, the central object is either a neutron star or a white dwarf. The gyro is any planetary or asteroid-like object orbiting the neutron star or a white dwarf. Moreover, the companion object that causes the tidal field can be a neutron star, white dwarf, a black hole, or a main-sequence star.
The tidal effect significantly affects the spacetime around the host star, which affects the gyro precession frequency. The gyro's precession frequency increases with the mass of the companion object and decreases as the separation between the host star and the companion star increases. The tidal effect also varies with the stiffness of the EoS of matter describing the NS.
We also find that the tidal field affects the spacetime around a white dwarf more than that of the neutron star.  \\

\end{abstract}

\section{Introduction}

The era of multimessenger astronomy has started with the simultaneous detection of gravitational waves, and their electromagnetic counterpart \cite{abbott}. The state of matter at high densities (finite chemical potential) has been an enigma for theoretical physicists for quite some time. The recent detection of gravitational event GW170817 from a binary neutron star merger (BNSM) and the precise measurement of NS's mass and radius hopes to put some constraints on the NS structure. \cite{annala,bauswein,Fattoyev,margalit,shibata,zhou,radice,rezolla}. 

Direct observation of neutron star (NS) cores is restricted: therefore, to know the matter properties at the star center, we need to model them. The nature of matter at such densities is still a challenge as there is no exact theory that describes strongly interacting matter at such densities (chemical potential). However, with more observation of events like GW170817 and exact measurement of NS mass and radius coming from Neutron Star Interior Composition Explorer (NICER) experiments, one expects to constrain the EoS of stellar matter \cite{nicer}. During the BNSM, in the inspiral phase, the star's tidal deformability (which depends on the EoS) is related to the quadrupole moment of the star and with the external tidal field. Before detecting GW170817, studies showed the star's tidal deformability would have a significant effect on the GW signals emitted during the inspiral phase \cite{hinderer,flanagan,hinderer2}. Analyzing the results of GW170817, the upper limit of the tidal deformability was obtained \cite{abbott}, which thereby constrained the EoS for the matter of NS interior.

The properties of strongly interacting matter at densities found in NS cores are difficult to model \cite{lattimer,brambilla}. On one hand the sign problem arises for lattice QCD calculation for non-zero chemical potential \cite{forcrand} while on the other perturbative QCD (pQCD) calculation \cite{kurkela} and effective model \cite{tews} becomes relevant at much higher densities. However, some robust properties on NS cores can be predicted from the constraints of the EoS at the low and high-density limits predicted from lattice QCD and pQCD \cite{kurkela2,fraga}. The uncertainties at the intermediate densities remain; however, with the recent observation of BNSM and NS mass and radius measurement, one expects to constrain them severely.

The late inspiral phase of BNSM generates strong tidal fields that deform the multipolar structure of the stars of the binary and its information imprinted on the gravitational wave waveform. The tidal deformability is closely related to the EoS of the stellar matter, and its detection puts severe constraints on the NS structure.

Their general relativistic effects define the spacetime around a massive astrophysical object like a black hole, NS, or a white dwarf. The properties around the central object can be captured by a test gyro, which rotates around in a fixed orbit outside the central objects. Previously, numerous studies try to capture the properties of the central object by observing the precession frequency of the gyro spinning outside the star \cite{lense,everitt,pugh,schiff,ciufolini,Iorio,Lucchesi,Renzetti,Will}. Studies relating to gyro precession around more exotic objects like black holes, NSs, and white dwarfs (WDs) has also been carried out in the recent past  \cite{hartle,glendenning,morsink,Chakraborty1,chatterjee}. In this article, we study the tidal deformability of the central object tidally deformed by an external tidal field (generated due to a companion). The gyro precession frequency can be seen as another astrophysical tool to study and limit NSs and WDs' tidal deformability. The gyro can be a planet orbiting around an NS or a WD.

The paper is arranged as follows: In section 2, we define our formalism for studying the GPF of a tidally deformed central object, and in section 3, we present our result for tidally deformed NSs and WDs. Finally, in section 4, we summarize our results and conclude from them.

\section{Formalism}

The tidal Love number measures the tidal effect on the star's orbital motion and the subsequent gravitational-wave signal \cite{love}. In Newtonian gravity \cite{solar}, the tidal
Love number is a constant of proportionality between the tidal field applied to the body and the resulting multipole moment of its mass distribution. For the general relativistic case, the method of finding quadrupole moment similar to the Newtonian way is no longer valid; however, the metric expansion holds in the star's local asymptotic rest frame \cite{thorne1998}. We start our calculation by studying a slowly rotating star tidally deformed by a static external gravitational field.\\

The metric for a spherically symmetric star in spherical polar coordinates is given by,
\begin{eqnarray}
ds^2\Big|_S=-e^{\nu(r)}dt^2+e^{\lambda(r)}dr^2+r^2d\theta^2+r^2\sin^2\theta d\phi^2
\end{eqnarray}
where the script '$S$' stands for spherical and $\nu(r)$ and $\lambda(r)$ are unknown functions of $r$ obtained numerically by solving the TOV equations \cite{tov}. In order to model a rotating star with rotational velocity $\Omega$, we incorporate the frame dragging frequency, $\omega(r)$. The metric describing the rotating star is given by,
\begin{align}
&ds^2\Big|_{\Omega}=-e^{\nu(r)}[1+2(h_0(r)+h_2(r)P_2(\cos\theta))]dt^2 \nonumber \\ 
&+e^{\lambda(r)}\Big[1+\frac{2e^{\lambda(r)}}{r}(m_0(r)+m_2(r)P_2(\cos\theta))\Big]dr^2 \nonumber  \\
&+r^2[1+2k_2(r)P_2(\cos\theta)]\Big(d\theta^2+\sin^2\theta (d\phi-\omega(r)dt)^2\Big)\nonumber\\
\end{align}
where, $h_0$, $m_0$, $h_2$, $m_2$, $k_2$ are unknown functions of $r$  \cite{hartle}. These unknown functions are perturbatively introduced using a dimensionless perturbation parameter,
\begin{eqnarray}
\epsilon_{\Omega}\sim\sqrt{\frac{\Omega^2R^3}{GM}}<1.
\label{epo}
\end{eqnarray}
The rotational velocity of the star $\Omega$ is of the order of $\sim O(\epsilon_{\Omega})$, thus the frame dragging frequency $\omega$ also is of the order of $\sim O(\epsilon_{\Omega})$. The perturbative functions is proportional to the centrifugal distortion of the star which goes as $\sim O(\Omega^2)\sim O(\epsilon_{\Omega}^2)$. The relative frame dragging frequencies is defined as $\bar{\omega}(r)\equiv\Omega-\omega(r)$, which is the net rotational frequency of the star that a freely falling body experiences.\\

\begin{figure}
	\includegraphics[width = 3.2in,height=1.6in]{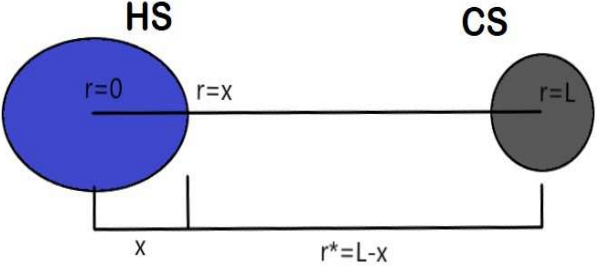}
	\caption{Tidal deformation on host star (neutron star or white dwarf) due to the presence of a companion star (neutron star or black hole)}
	\label{fig1}
\end{figure}

In the presence of a Companion Star (CS) (which can be a BH, NS or a WD), the host NS or WD experiences tidal deformations. We name the tidaly deformed NS or WD as the host star (HS). This tidal deformation is induced on the HS due to the difference in gravitational field $g$ of the CS at $r=0$ and $r=x$ as shown in fig \ref{fig1}. This difference can be written as difference in derivative of the potential $\Phi$ as $g(x)-g(0)=-\partial_i\Phi|_{r=x}+\partial_i\Phi|_{r=0}$, where $g(x)$ is the gravitational field at any point $r=x$. Taylor expanding $g(x)$ around $r=0$, we have
\begin{equation}
g(x)=g(0)-\Big[\partial_j\partial_i\Phi|_{r=0}\Big]x^j+O(x^2)
\end{equation}
and upto first order in $x^j$, and the tidal field $E_{ij}$ is defined as
\begin{eqnarray}
E_{ij}\equiv\frac{\partial^2\Phi}{\partial x^i\partial x^j}.
\end{eqnarray}
For calculating $\Phi$ for CS at some point $x$, we have used the coordinate $r*=L-r$ as shown in fig \ref{fig1}. The quadrapole moment $Q_{ij}$ generated on the HS is proportional to the tidal field of the CS, $Q_{ij}\propto E_{ij}$, or upto order $E_{ij}\sim O(\epsilon_T)$ where
\begin{equation}
\epsilon_T\sim\sqrt{\frac{\Phi R}{GM}}<1 .
\end{equation}
One can write
\begin{eqnarray}
Q_{ij}=-\lambda E_{ij},
\label{love}
\end{eqnarray}
the proportionality constant $\lambda$ is the tidal love number. $\lambda$ depends on the properties of the HS such as Equation of State (EoS), central density etc. The metric for HS with such tidal deformation having an induced quadrapole moment $Q_{ij}$ and is written as
\begin{eqnarray}
ds^2\Big|_T=g^{S}_{\mu\nu}+g^T_{\mu\nu}.
\end{eqnarray}
The metric $g^S_{\mu\nu}$ is our spherically symmetric metric and the tidal metric is given by
\begin{align}
g^T_{\mu\nu}=diag\Big[-e^{\nu(r)}H_0(r),e^{\lambda(r)}H_2(r), \nonumber\\
r^2K(r),r^2\sin^2\theta K(r)\Big]Y_{2m}(\theta,\phi)
\end{align}
where the script '$T$' stands for tidal and $Y_{2m}$ are the spherical harmonics. The unknown functions of $H_0$, $H_2$, $K$ are of the order $\sim O(\epsilon_T)$. The spherical harmonics are given by
\begin{eqnarray}
Y_{2m}=\sqrt{\frac{5}{4\pi}\frac{(2-m)!}{(2+m)!}}e^{im\phi}P_2^m(\cos\theta).
\end{eqnarray}
From equation \ref{love}, the expansion of $Q_{ij}$ and $E_{ij}$ gives
\begin{align}
\sum_{m=-2}^{m=2}Q_mY^{2m}_{ij}=-\lambda\sum_{m=-2}^{m=2}E_mY^{2m}_{ij}
\end{align}
where $Y_{2m}\equiv Y^{2m}_{ij}n^in^j$ and\\ $n^i\equiv(\sin\theta\cos\phi,\sin\theta\sin\phi,\cos\theta)$. Thus
$Q_m=-\lambda E_m$ and any one choice of $m$ is enough for solving for the tidal love number $\lambda$ \cite{hinderer2}. To retain axisymmetry we take $m=0$ mode and the metric becomes
\begin{align}
g^T_{\mu\nu}=diag\Big[-e^{\nu(r)}H_0(r),e^{\lambda(r)}H_2(r),\nonumber\\
r^2K(r),r^2\sin^2\theta K(r)\Big]P_2(\cos\theta).
\end{align}
\subsection{Tidally deformed rotating star}
The tidally deformed metric for a rotating star becomes
\begin{eqnarray}
g_{\mu\nu}=g^S_{\mu\nu}+g^T_{\mu\nu}+g^{\Omega}_{\mu\nu}.
\end{eqnarray}
We solve for equation of motion from the Einstein equation for the order $O(\epsilon_{\Omega})$ and $O(\epsilon_T)$ where the order $O(\epsilon_{\Omega})$ gives the Hartle-Throne equation for frame dragging and $O(\epsilon_{T})$ the tidal equation of motion respectively. Hence in order to find the coupling between rotation and tidal deformation, we must solve the Einstein equation upto $O(\epsilon_{\Omega}\epsilon_T)$. To obtain the order $O(\epsilon_{\Omega}\epsilon_{T})$, we expand the frame dragging frequency $\omega$ in powers of $\epsilon_T$ as
\begin{eqnarray}
\omega(r)=W^0(r)+W^1(r)+W^2(r)+W^3(r).........
\end{eqnarray} 
such that, $W^0\sim O(\epsilon_{\Omega})$, $W^1\sim O(\epsilon_{\Omega}\epsilon_T)$, $W^2\sim O(\epsilon_{\Omega}\epsilon_T^2)$ and so on. Thus redefining $\omega\equiv W^0$ and $W\equiv W^1$, we solve the Einstein equation $G_{\mu\nu}=8\pi T_{\mu\nu}$ upto $O(\epsilon_{\Omega}\epsilon_T)$ to get the coupled frame dragging differential equation. The coupled frame dragging equation is given by
\begin{align}
&\frac{d^2W}{dr^2}+F_1(r)\frac{dW}{dr}+F_2(r)W=F_3(r,\theta)\Big(S_1(r,\theta)\nonumber\\
&+S_2(r,\theta)+S_3(r,\theta)+S_4(r,\theta)-S_5(r,\theta)+S_6(r,\theta)\Big)
\label{eqW}
\end{align}
where the unknown coefficients $F_1$, $F_2$, $F_3$ are given by
\begin{align}
&F_1(r)\equiv-\frac{\left(r \left(\lambda '(r)+\nu '(r)\right)-8\right)}{2 r}\\
&F_2(r)\equiv-\frac{2e^{-\frac{1}{2} (\lambda (r)+\nu (r))}}{8 \sqrt{\pi } r^2 P'} \Big(64 \pi ^{3/2} r^2 P' \rho (r) e^{\frac{1}{2} (3 \lambda (r)+\nu (r))}\nonumber\\
&+8 \sqrt{\pi } \left(8 \pi  r^2+1\right) P(r) P' e^{\frac{1}{2} (3 \lambda (r)+\nu (r))}\Big)\\
&F_3(r,\theta)\equiv\frac{\sin^2(\theta ) e^{\frac{1}{2} (-3) (\lambda (r)+\nu (r))}}{8 \sqrt{\pi } P'}
\end{align}
\\
The source terms from tidal deformation ($S_1$, $S_2$, $S_3$, $S_4$, $S_5$ and $S_6$) are given by,\\
\begin{align}
&S_1(r,\theta)\equiv -4 \sqrt{5} P' \omega (r) e^{\frac{1}{2} (3 \lambda (r)+\nu (r))} \Big(P_2(\cos (\theta )) \Big(3 (K[r]\nonumber\\
& +H(r))-Mass(r) H'(r)\Big)+\cos (\theta ) K[r] P_2'(\cos (\theta ))\Big)\\
&S_2(r,\theta)\equiv 8 \pi  \sqrt{5} r^2 P' \rho (r) P_2(\cos (\theta )) e^{\frac{1}{2} (3 \lambda (r)+\nu (r))} \nonumber\\
&\Big(-4 \Omega  K[r]-r \omega (r) H'(r)+H(r) (\omega (r)+4 \Omega )\Big)\\\nonumber\\
&S_3(r,\theta)\equiv 4 \sqrt{5} P(r) P' P_2(\cos (\theta )) e^{\frac{1}{2} (3 \lambda (r)+\nu (r))} \nonumber\\
&\Big(2 \pi  r^2 \Big(12 \omega (r) K[r]-8 \Omega  K[r]+r \omega (r) H'(r)\nonumber\\
&+5 H(r) \omega (r)+4 \Omega  H(r)\Big)-H(r) \omega (r)\Big)\\\nonumber\\
&S_4(r,\theta)\equiv 8 \pi  \sqrt{5} r^2 \rho ' P_2(\cos (\theta )) e^{\frac{1}{2} (3 \lambda (r)+\nu (r))}\nonumber\\
& (4 P(r) (\omega (r)-\Omega )K[r]+H(r) \omega (r) (P(r)+\rho (r)))\\\nonumber\\
&S_5(r,\theta)\equiv \sqrt{5} P' \omega (r) P_2(\cos (\theta )) e^{\frac{1}{2} (\lambda (r)+\nu (r))} \nonumber\\
&\Big(r H'(r) \left(-r \lambda '(r)+r \nu '(r)+4\right)+2 H(r)\nonumber\\
&\left(r \nu '(r) \left(r \nu '(r)+2\right)-4\right)\Big)\\\nonumber\\
&S_6(r,\theta)\equiv 4 \sqrt{5} r H(r) P' P_2(\cos (\theta )) (\Omega -\omega (r))\nonumber\\ &\left(\lambda '(r)+\nu '(r)\right)
\end{align}
Equation \ref{eqW} is then solved numerically by simultaneously solving EOM for rotation and EOM for tidal deformation upto $O(\epsilon_{\Omega})$ and $O(\epsilon_{T})$ respectively.
\subsection{Tidal love number}
The tidal love number $\lambda$ is calculated by first redefining $\lambda$ as $k_2$ to make it dimensionless
in geometrized units ($G=c=1$) as
\begin{eqnarray}
k_2\equiv\frac{3}{2}\lambda R^{-5}
\end{eqnarray}
where $R$ is the TOV radius. The expression for $k_2$ is given by \cite{hinderer2}
\begin{align}
&k_2=\frac{8C^2}{5}(1-2C^2)[2+2C(y-1)-y]\Bigg[2C(6-3y\nonumber\\
&+3C(5y-8))+4C^3(13-11y+C(3y-2)+2C^2(1+y))\nonumber\\
&+3(1-2C^2)[2-y+2C(y-1)]ln(1-2C)\Bigg]^{-1}
\end{align}
where  $C\equiv M/R$ is the ratio of star's TOV mass and radius. The quantity $y$ is obtained from the differential equation
\begin{align}
&r\frac{dy}{dr}+y(r)^2+y(r)e^{\lambda(r)}\Big[1+4\pi r^2[P(r)\nonumber\\
&-\rho(r)]\Big]+r^2Q(r)=0
\label{diff-eqn}
\end{align}
where $y$ is defined as $y\equiv \Big|r\frac{dH(r)}{dr}/H(r)\Big|_{r=R}$ and $Q(r)$ is a known function of $r$. Equation \ref{diff-eqn} is solved with appropriate boundary condition to calculate $k_2$.

\subsection{calculation of overall precession frequency}
Solving the equation \ref{eqW} numerically and obtaining other metric coefficients by solving TOV equations, one obtains the details of the ST (Spacetime) in and around the HS, which is tidally deformed by an external tidal field caused by the CS. The ST's details can be captured by a gyro orbiting around the HS \cite{kknrtm}. To study the tidal effect on the GPF (Gyro precession frequency) due to the presence of a companion object, we have to compare the tidally perturbed case with the case of no companion. The unperturbed case with gyro orbiting around an isolated NS was done previously \cite{kknrtm}. In this paper, we find the amount of change the tidal field of companion object will make in the GPF moving around an HS.\\
In a stationary and axisymmetric ST with coordinates $ t, r, \theta, \phi $, GPF becomes \cite{Chakraborty:2016mhx}:
\begin{eqnarray}
\vec{\Omega}_P&=&\frac{1}{2\sqrt {-g}\left(1+2\Omega\frac{g_{t\phi}}{g_{tt}}+\Omega^2\frac{g_{\phi\phi}}{g_{tt}}\right)}. \nonumber
\\
&&\hspace{-5mm} \left[-\sqrt{g_{rr}}\left[\left(g_{t\phi,\theta}
-\frac{g_{t\phi}}{g_{tt}} g_{tt,\theta}\right)+\Omega\left(g_{\phi\phi,\theta}
-\frac{g_{\phi\phi}}{g_{tt}} g_{tt,\theta}\right) \right.\right. \nonumber
\\
&&\hspace{-5mm} \left.\left.+ \Omega^2 \left(\frac{g_{t\phi}}{g_{tt}}g_{\phi\phi,\theta}
-\frac{g_{\phi\phi}}{g_{tt}} g_{t\phi,\theta}\right) \right]\hat{r} \right. \nonumber
\\
&&\hspace{-5mm}\left. +\sqrt{g_{\theta\theta}}\left[\left(g_{t\phi,r}
-\frac{g_{t\phi}}{g_{tt}} g_{tt,r}\right) +\Omega\left(g_{\phi\phi,r}
-\frac{g_{\phi\phi}}{g_{tt}} g_{tt,r}\right) \right.\right. \nonumber
\\
&&\hspace{-5mm}\left.\left.+  \Omega^2 \left(\frac{g_{t\phi}}{g_{tt}}g_{\phi\phi,r}
-\frac{g_{\phi\phi}}{g_{tt}} g_{t\phi,r}\right) \right]\hat{\theta} \right] .
\label{main}
\end{eqnarray}

\subsection{Circular Geodesics}

The GPF is now calculated for the gyroscopes revolving along circular geodesics around the HS. The angular velocity (denoted as $\Omega=\Omega_g$) of a particle moving on a circular equatorial geodesic is:
\begin{equation}
\Omega_g = \frac{d\phi}{dt} = \frac{-\partial_{r} g_{t\phi} \pm \sqrt{(\partial_{r}g_{t\phi})^2 - \partial_{r}g_{tt}\partial_{r}g_{\phi\phi}}}{\partial_{r}g_{rr}}.
\end{equation} 

In the presence of a tidal field the orbital angular velocity $\Omega$ is shifted which is given by 
\begin{align}
&\Omega_{gt} = \frac{-4 \sqrt{\frac{1}{16} r \xi \left(e^{\nu (r)} \xi_1 +8 r \xi_2 (W(r)+\omega (r))  \right)+\xi_3^2}-\xi_4}{r \xi}
\end{align}

where  \\
$\xi$ = $\left(\sqrt{\frac{5}{\pi }} \left(2 K[r]+r K'(r)\right)-8\right)$\\
$\xi_1$ = $\left(\sqrt{\frac{5}{\pi }}
\left(H'(r)+H(r) \nu '(r)\right)-4 \nu '(r)\right)$\\
$\xi_2$ = $\left(r \left(W'(r)+\omega '(r)\right)+W(r)+\omega
(r)\right)$\\
$\xi_3$ = $\left(r^2 \left(W'(r)+\omega '(r)\right)+2 r (W(r)+\omega (r))\right)$\\
$\xi_4$ = $4 r \left(r \left(W'(r)+\omega '(r)\right)+2 W(r)+2
\omega (r)\right)$.
\section{Results}
	
\begin{figure}
\centering
\includegraphics[width = 3.4in,height=1.2in]{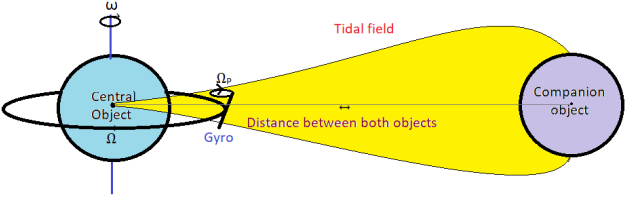}
\caption{The sketch shows the qualitative strength of the tidal field caused by companion object and it's effect on the central object. Here the HS is a rotating NS. The CS could be a NS or BH. }
\label{sk}
\end{figure}	

This section deals with the numerical calculation of the GPF around a tidally disrupted HS caused by a CS. The precession frequency is studied in a region of space around the HS where $\omega > W$ for the perturbative approach to be valid ($\omega$ is the frame-dragging frequency and $W$ is the correction to frame dragging frequency due to tidal effect). The amount of separation between the HS and CS in most of the cases is taken to be $\sim$ $0.1$ AU (when the HS is a Neutron star). Any separation more than this distance will also remain valid for the perturbative approach to hold for the gyro orbits we have chosen, which will be seen in the subsequent section. 

Figure \ref{sk} shows a sketch of the qualitative nature of our problem. It shows how a tidal field caused by a companion object affects the central or the host star. The gyro's position is also depicted, which is kept between the CS and HS and nearer to the HS. This is done as the gyro primarily studies the tidal field's effect on the ST of HS. 
	
\subsection{Neutron Stars}
	
\begin{figure}
\centering
\includegraphics[width = 3.4in,height=2.5in]{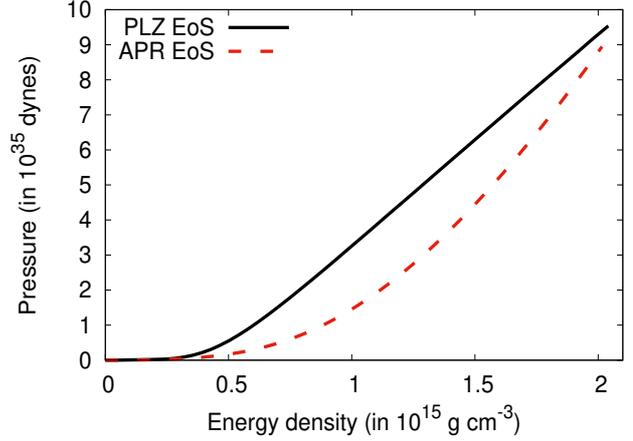}
\caption{The figure shows the nature of the two tabulated EoS used in this article. The plot in black is for PLZ EoS and the red dotted curve is for APR EoS. }
\label{f23}
\end{figure}

Let us first consider that the HS is an NS. Figure \ref{f23} shows the nature of the EoS used in this article to model an NS structure. The EoS which we have considered to describe the matter properties (APR and PLZ) satisfies the recent constraints of max mass bound $M_{max}$ $\geq$ $1.97$ and tidal deformability $\frac{\lambda}{M^5}$ $\leq$ 580. The PLZ EoS is the stiffer one, while the APR EoS is softer than it. The mass of the NS which forms the HS is taken to be $1.4$ solar mass.

\begin{figure}
\centering
\includegraphics[width = 3.4in,height=2.5in]{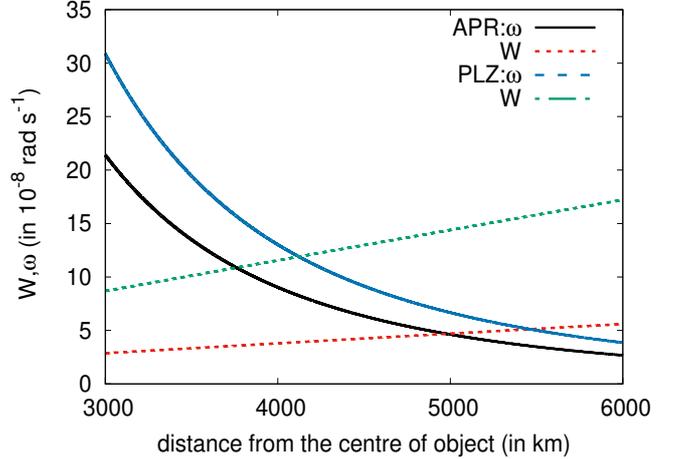}
\caption{Here the variation of $\omega$ and $W$ are shown with respect to distance in the exterior region around the HS. The separation between the two objects is kept as $1\times10^{7}$ km. The mass of the companion is 1.0 $M_o$, where $M_o$ is the solar mass. The black solid line and red dotted line indicates the plot for APR EoS while the blue dashed and green dot dashed line are for PLZ EoS. }
\label{f5}
\end{figure}

\begin{figure}
\centering
\includegraphics[width = 3.4in,height=2.5in]{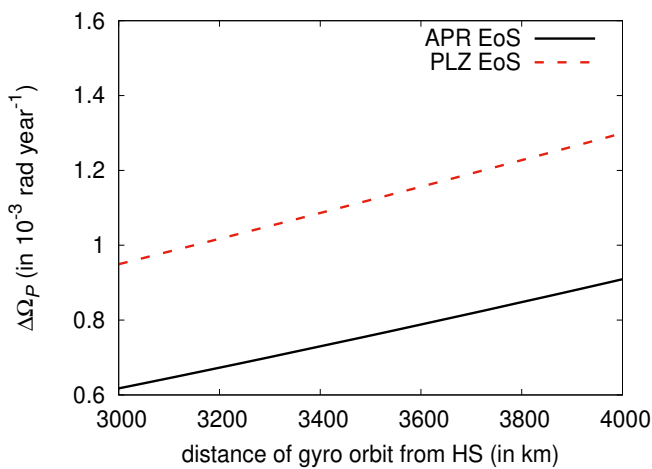}
\caption{Variation of change in GPF with distance outside the HS. The separation between the two objects is kept as $1\times10^{7}$ km. The mass of the CS is 1.4 $M_o$ and mass of the HS is 1 $M_o$. The black solid line is for the HS model with APR EoS while the red dotted line is for NS with PLZ EoS. }
\label{f6}
\end{figure}

Our calculation for the GPF is perturbative, and it has a range of validity. The range of validity for the gyro frequency is the range of distance around the HS where $\omega$ $>$ $W$. Therefore, for the calculation to be valid, we must keep the gyro orbit at a distance from HS where $\omega>W$. The figure \ref{f5} shows the range of validity for an HS, which is a 1.4 $M_o$ NS, where $M_o$ is the solar mass. The value of $W$ becomes greater than the value of $\omega$ after reaching a certain distance (limiting distance $L_l$) as the tidal field strength increases as one goes away from the HS and towards the CS. Beyond $L_l$, the perturbative approach is invalid. The $L_l$ distance depends on the EoS of the HS, the mass of CS, and the separation between them. The above figure is drawn with the separation between the objects being $1\times10^7$ km and the mass of the CS being 1 $M_o$. The limiting distance is farther for APR EoS than that of the PLZ EoS. The value of $L_l$ for CS with APR EoS is around $5000$ km while that for PLZ EoS is around $4200$ km. If the companion star is at a distance $\sim$ 1 AU, then the limiting distance $L_l$ shifts to $\sim$ 20000 km for both EoS. Pulsar planets or massive asteroid-like objects within this range can serve as a test gyro for studying the GPF. The change in their GPF can be examined to detect any tidal field caused by a CS, along with the information about HS's properties. In figure \ref{f5} we can see that, in the case of PLZ EoS the value of both $\omega$ and $W$ are greater than that of APR EoS. That is why the amount of change in the GPF due to tidal effect is greater in the case of PLZ EoS, which will be seen subsequently.  

The change in the GPF is measured in terms of a quantity $\Delta \Omega_P$. This quantity is the difference between the GPF around the HS with a companion and without a companion. $\Delta \Omega_P$ is defined as: 
\begin{equation}
\Delta \Omega_P = \lvert\Omega_P - \Omega_{Pt}\rvert
\end{equation}  
where $\Omega_P$ is the overall precession frequency of gyro orbiting around the HS but without a companion, while $\Omega_{Pt}$ is the GPF in the case where HS has a companion. $\Omega_P$ and $\Omega_{Pt}$ are in $rad$ $year^{-1}$. 

In figure \ref{f6}, we show the change in precession frequency with an increase in the size of the gyro's orbit as it moves away from HS (towards the CS). Within the range of validity of the perturbative approach, the tidal strength keeps increasing, and hence the tidal field contributes more to the coupling than dragging. This, too, is reflected in the change in precession frequency. The range of distance (3000-4000 km) shown in figure \ref{f6} is to show the maximum impact of the tidal field within the range of validity. This range of distance is valid for the NS model with both the cases of EoS of matter. The perturbative approach is no longer accurate after a distance of $\sim$ 4200 km for an NS with PLZ EoS and after a distance of $\sim$ 5000 km for an NS with APR EoS. This is for the case where the CS is at a distance of $\sim$ $10^7$ km from HS. If this separation is increased, the limiting distance of $L_l$ also increases.    

\subsubsection{Change in the mass of companion}

\begin{figure}
\centering
\includegraphics[width = 3.4in,height=2.5in]{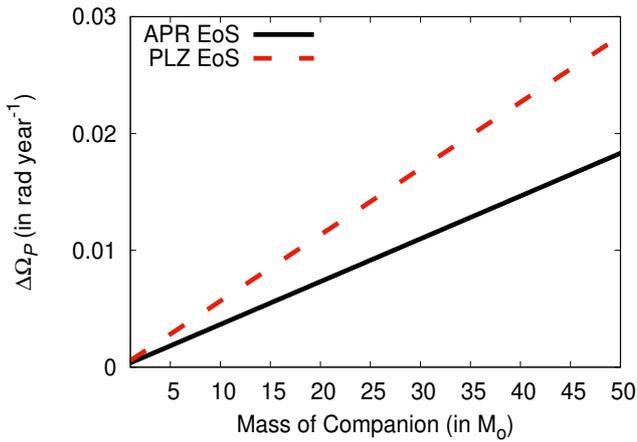}
\caption{Variation of change in GPF with increasing mass of companion object. The gyro is kept at a distance of $2000$ km from the HS centre. Separation is $1\times10^{7}$ km and the mass range is 1$M_o$ - 50$M_o$, where $M_o$ is the solar mass. The black solid line is for the case with APR EoS. And the red dotted line is for the HS model with PLZ EoS.}
\label{f3}
\end{figure}

As stated earlier, the GPF depends on the mass of the CS. In figure \ref{f3}, we show how the GPF changes with the change in the CS mass. The HS's mass is 1.4 solar mass, and the distance between the HS and CS is $10^7$ km.  
The radius of the orbit of the gyro is at $\sim$ $2000$ km. This choice of $2000$ km is taken based on two key points. Firstly,  it lies in the outer region and sufficiently far from the star's surface but within the perturbative range of validity for CS of mass up to 50 $M_o$. Secondly, it is close to the limiting distance of validity, where the tidal effect is maximum within the allowed region. 

\begin{figure}
\centering
\includegraphics[width = 3.4in,height=2.5in]{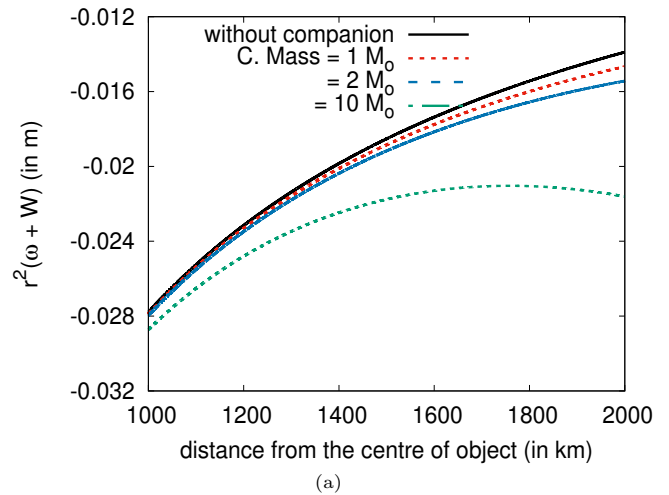}
\scriptsize{(a)}
\includegraphics[width = 3.4in,height=2.5in]{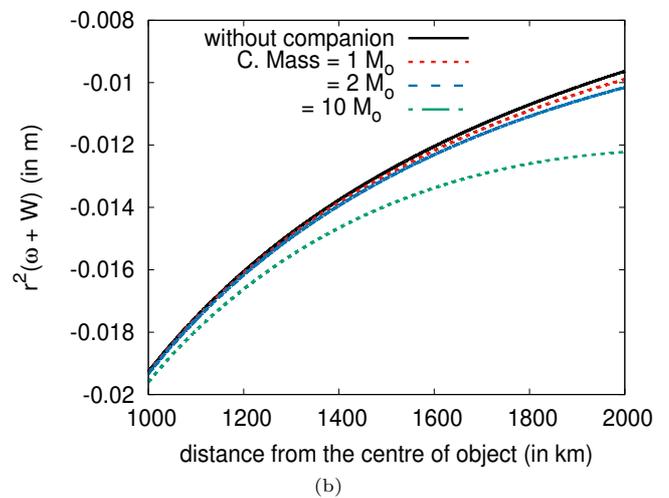}
\scriptsize{(b)} 
\caption{Variation of frame-dragging effect with distance from the centre of HS. Separation is $1\times10^{7}$ km and the mass range is 1$M$ - 10$M$, where $M$ is the solar mass. (a)HS model with PLZ EoS. The black solid line is for the case without companion and the red dotted line is for the case with companion mass 1 $M_o$ and the blue-dashed line is for the case with comapnion mass of 2 $M_o$. The green dot dashed line is for CS of 10 $M_o$. (b)HS model with APR EoS. Nomenclature remains same.}
\label{f21}
\end{figure}

Figure \ref{f3} shows that the more massive the companion, the greater is the change caused in the GPF. Even with the same mass of the HS, the change in the GPF differs with different EoS. The rate of change of $\omega$ and $W$ are different for PLZ and APR EoS, showing a different amount of changes in precession frequency for the same mass of the companion. This change in the GPF with and without a CS can be analyzed by studying figure \ref{f21}. For PLZ EoS, the change in the dragging strength due to CS of same mass is more (figure \ref{f21} (a)), than that of APR EoS (figure \ref{f21} (b)). The dragging strength is the frame-dragging coefficient in the metric of a given rotating spacetime. In this case, as we have considered gyro orbits along the equatorial plane, therefore, $\theta=\frac{\pi}{2}$. Thus the coefficient is $r^2(\omega+W)$. The higher the change in the Frame dragging coefficient, the greater the change in GPF. The CS of 1 $M_o$ is probably a WD (or even a small NS) or a main-sequence star. The CS of 2 $M_o$ can be a main-sequence star or an NS. Moreover, the CS of 50 $M_o$ can be thought to be a stellar BH. 	

\subsubsection{Change in the amount of separation}

\begin{figure}
\centering
\includegraphics[width = 3.4in,height=2.5in]{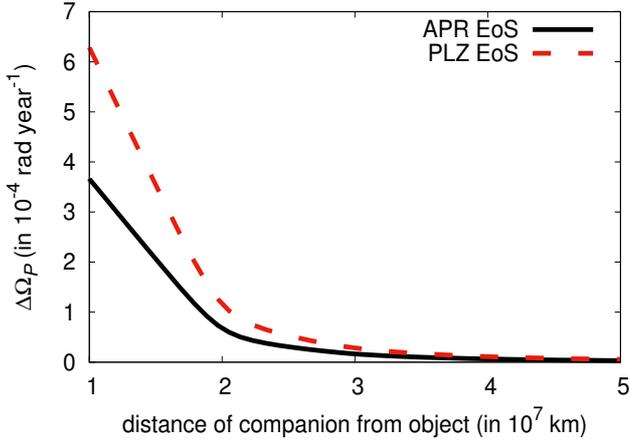}

\caption{Variation of change in GPF at a distance outside the HS with respect to the change in separation between the two objects. The range of separation is $1\times10^{7}$ km - $5\times10^{7}$ km. The solid black line is for the HS model with APR EoS while the red dotted line is for the HS model with PLZ EoS.}
\label{f1}
\end{figure}

The change in the GPF also depends on the distance between the HS and CS. In figure \ref{f1}, we show how the GPF varies with the distance between HS and CS. The figure is plotted with both the HS and the CS having the same mass (1.4 solar mass), and the distance between the two-star varies between the range of $1\times10^{7}$ km - $5\times10^{7}$ km. The distance range is chosen so that the change in GPF is prominent. The separation between two objects in case of them being NS-NS, NS-WD, or NS-BH binaries observed is being taken into consideration. If the separation is more than the change in GPF eventually become less.

Figure \ref{f1} shows that increasing the separation between HS and CS decreases the value of precession frequency because of the decrease in tidal strength value $W$ (for a particular radius of gyro orbit). The distance of the gyro orbit is at $2000$ km from the center of HS. The change in the GPF with increasing separation is more for the PLZ case than that of APR. The decrease of tidal strength coupled to frame dragging will be more for PLZ (with increasing separation between CS and HS) than APR, which we can see from figure \ref{f21}.  

\subsection{White Dwarf}

The HS need not always be an NS, and it can even be a WD. 
In this section, we study companion objects' tidal effect on the precession frequency of a gyro orbiting around a WD. While constructing the WD, we have considered relativistic polytropic EoS for describing the matter inside a WD. \\

Parameters used for constructing a WD\\
Polytropic index: $\gamma$=$\frac{4}{3}$\\
central density = $3.0\times10^6$ $gcm^{-3}$\\
mass = $1.0$ $M_o$, where $M_o$ is the solar mass\\      
radius =  $6400$ $km$   \\   
The above configurations are for White Dwarfs resembling Sirius-B. This WD of 1 $M_o$ can be considered as a fairly massive WD.

\begin{figure}
\centering
\includegraphics[width = 3.4in,height=2.5in]{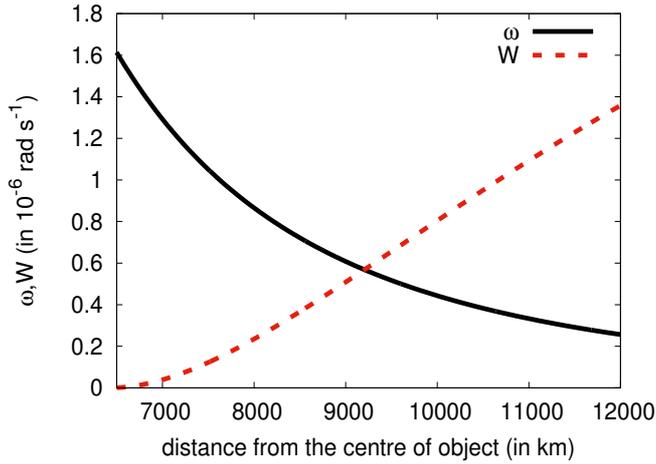}
\caption{Variation of frame dragging frequency $\omega$ and tidal correction term $W$ with distance from the HS. The separation between the two objects is kept as $1\times10^{10}$ km. The mass of the CS is 1 $M_o$. The black line shows the frame dragging frequency $\omega$ while the red dotted line shows the tidal correction $W$. }
\label{f13}
\end{figure}

As the WD is less compact than an NS, the perturbation limit range is different for WDs.
From Figure \ref{f13} we can see that the perturbation limit $L_l$ is around $9300$ km. This is the validity range when the HS and CS distance is $10^{10}$ km and for a CS mass of 1 $M_o$. With this configuration, the gyro could be anything that orbits the WD at a maximum distance of $9300$ km. 

\begin{figure}
\centering
\includegraphics[width = 3.4in,height=2.5in]{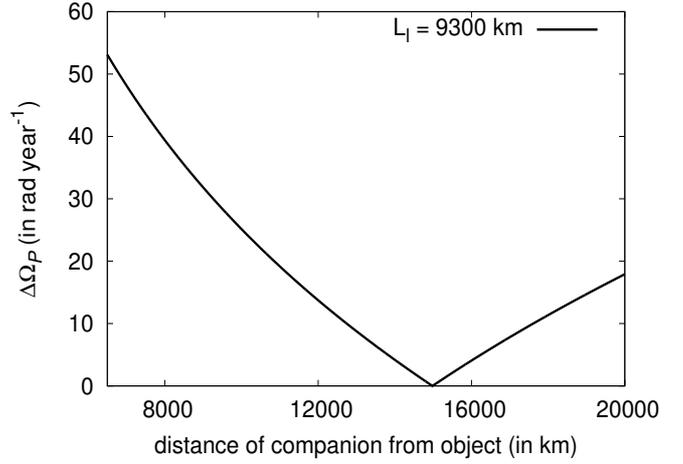}
\caption{Variation of $\Delta \Omega_P$ with distance outside the HS. The separation between the two objects is kept as $1\times10^{10}$ km. The mass of the CS is 1 $M_o$. The black solid line is for the case with companion and the red dotted line is for the case without companion. }
\label{f14}
\end{figure}

As the WD is less compact, the tidal effect on the WD is much more prominent, which can be seen from figure \ref{f14}. The order of change in GPF is high comparing the value of $\Delta \Omega_P$ around WD with that of NS. Close to the surface, the value of $\omega$ is high, and thus $\Omega_{P}$ is higher than $\Omega_Pt$. However, $\Omega_{Pt}$ rises above the value of $\Omega_P$ after reaching around 15000 km. But this distance is beyond the limiting distance $L_l=9300$ $km$. Up to $\sim$ 15000 km, the drag contributes more towards the coupling of tidal effect and frame-dragging. Further than this distance, the tidal correction contributes more towards coupling.

\begin{figure}
\centering
\includegraphics[width = 3.4in,height=2.5in]{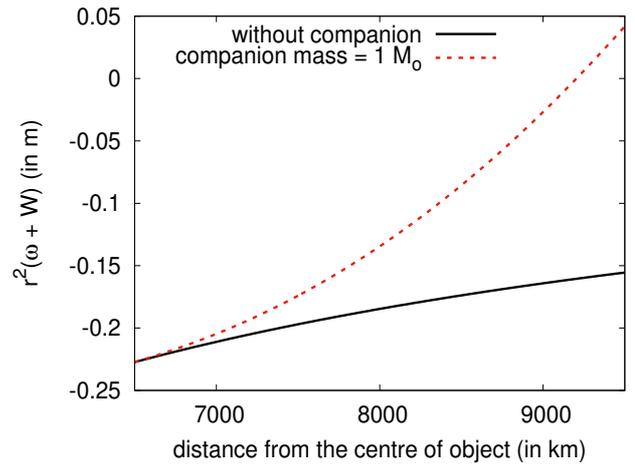}
\caption{Variation of frame dragging term with distance from HS. The separation between the two objects is kept as $1\times10^{10}$ km. The mass of the CS is 1 $M_o$. The black solid line is for the case without companion and the red dotted line is for the case with companion of 1 $M_o$.}
\label{f22}
\end{figure}

Figure \ref{f22} shows that the tidal correction to the metric's frame-dragging coefficient around a realistic WD. The solid black line is for the case without a companion, and the red dotted line is for the case with the companion of 1 $M_o$. The increase in the dragging coefficient's value is more in the case with a companion as it includes non-zero tidal correction $(W)$. If a lower mass WD is considered, then $L_l$ shifts towards the surface of HS. This will eventually restrict our range of validity.

\section{Summary and Conclusion}

In this work, we have studied how the GPF can be used to study the companion object's tidal effect on the spacetime of HS. The companion object can be an NS, a WD, or a stellar BH. The gyro can be a planet or an asteroid like object revolving around an NS or a WD along the circular geodesic. The tidal field significantly affects the ST around an HS, and a gyro orbiting the HS in a circular orbit can give us information about the tidal effect. In the presence of a companion object, the tidal effect can deform the HS. This deformation causes a change in the ST and can subsequently change the GPF of the gyro orbiting the HS. From this change, the properties of HS and the mass of CS can be identified qualitatively.

The modeling of the HS is done, assuming slow rotation. The host object's rotation frequency is taken small so that the rotation does not contribute significantly to the deformation. That is why perturbation to the dragging is taken up to first order in $\epsilon_{\Omega}$. The orbit of the CS and HS revolving around each other is taken to be circular as an approximation to retain axisymmetry in the perturbated metric.

We have studied the case of HS being an NS and a WD.  In NS's case, we have shown that increasing the mass of the CS makes the tidal correction to drag greater and hence causing a more significant change to the GPF. Increasing the separation between CS and HS decreases the tidal strength at a given gyro orbital distance and hence decreases the value of GPF. The drag and the tidal correction vary differently with distance from the center of HS with varying EoS of matter, which reflects in the change of GPF.
In WD's case, the GPF is affected considerably due to softness in the EoS of matter. WD matter is much less compact than that of NS. In WD's case, the tidal effect due to companion of the same mass and same distance affects the spacetime much more than that of NS. 

The perturbative approach used here turns out to be limited as there always exists a limiting distance beyond which we cannot place our gyro. However, within this range, if the precession frequency of planetary objects or asteroid-like objects orbiting around pulsars or WD can be measured, the difference in GPF can be seen in an isolated object and the case of binaries.
The GPF can become a tool to identify tidal effects for objects in a binary. Our present endeavor is towards developing a more exact formalism (instead of a perturbative approach) by which one can study the tidal effects and calculate GPF around an object.

\section{Acknowledgments}
The authors are grateful to IISER Bhopal for providing all the research and infrastructure facilities. RM would also like to thank the SERB, Govt. of India, for monetary support in the form of Ramanujan Fellowship (SB/S2/RJN-061/2015) and Early Career Research Award (ECR/2016/000161). 

\section*{Data Availability Statement} This manuscript has no associated data or
the data will not be deposited. [Authors comment: This is an theoretical
work and there is no additional data.]

\end{document}